\pdfoutput=1
\documentclass[12pt]{article}
\usepackage{cite}
\usepackage{graphicx}
\usepackage{latexsym}
\usepackage{mathrsfs}
\usepackage[overload]{textcase}

\font\grande=cmr9.5 scaled \magstep4
\font\medio=cmr9.5 scaled \magstep2
\outer\def\beginsection#1\par{\medbreak\bigskip
      \message{#1}\leftline{\bf#1}\nobreak\medskip
\vskip-\parskip
      \noindent}

\begin{document}
\bibliographystyle{unsrt}

\titlepage

\vspace{1cm}
\begin{center}
{\grande The maximal frequency of cosmic gravitons}\\
\vspace{1.5 cm}
Massimo Giovannini\footnote{e-mail address: massimo.giovannini@cern.ch}\\
\vspace{1cm}
{{\sl INFN, Section of Milan-Bicocca, 20126 Milan, Italy}}
\vspace*{1cm}
\end{center}
\vskip 0.3cm
\centerline{\medio  Abstract}
\vskip 0.5cm
We show that the maximal frequency of cosmic gravitons must not exceed the THz domain. 
From a classical viewpoint, both in conventional inflationary scenarios and in bouncing models the largest frequency of the spectrum overshoots the MHz band even if its specific signature is model dependent. According to a quantum mechanical perspective the maximal frequency  is instead associated with the range of energies where a single pair of gravitons with opposite (comoving) three-momenta is produced. The upper limit on the largest frequency determines the minimal chirp amplitude [typically ${\mathcal O}(10^{-32})$] required for a direct detection of a cosmic signal in the THz band. Below this limiting frequency the minimal chirp amplitude can be enhanced so that the optimal range ultimately depends on the physical properties of the diffuse backgrounds. In case a hypothetical instrument (at present just a figment of a hopeful imagination) would reach chirp amplitudes down to ${\mathcal O}(10^{-30})$ in the MHz or GHz bands, the Bose-Einstein correlations could be used to probe the properties of cosmic gravitons and their super-Poissonian statistics. 
\noindent
\vspace{5mm}
\vfill
\newpage
Three classes of direct bounds are today constraining the cosmic gravitons in a group of complementary frequency windows ranging between the aHz domain and the kHz band. In the aHz region (we recall $1\,\, \mathrm{aHz} = 10^{-18}\,\, \mathrm{Hz}$) the direct measurements of the temperature and polarization anisotropies set relevant limits on the tensor-to-scalar ratio $r_{T}$ so that, according to current data, $r_{T} < 0.06$ or even $r_{T} <0.03$ \cite{AD1,AD2}. In the nHz window (we recall $1\,\, \mathrm{nHz} = 10^{-9} \,\, \mathrm{Hz}$) the pulsar timing arrays report evidences potentially associated with diffuse backgrounds of gravitational radiation \cite{PT1,PT2} even if the competing experimental collaborations still make different statements about the correlation properties of the observed signal which should eventually comply, in the case of relic gravitons, with the Hellings-Downs curve \cite{PT3}. Finally in the audio band (between few Hz and $10$ kHz) the operating wide-band detectors set an upper bound on the spectral energy density of relic gravitons for a handful of spectral slopes \cite{INT1,INT2}. As we move from the small frequency domain to the high-frequency region the astrophysical signals progressively disappear until the MHz region where the spectrum is plausibly dominated by cosmic gravitons, i.e. particles produced by the early variation of space-time curvature \cite{GR1,GR2}. The question addressed in this investigation is, in short, the following: what is the largest frequency of cosmic gravitons? The answer to this question may have both theoretical and observational implications especially because it seems rather awkward to imagine a source of high-frequency gravitational radiation more efficient than the early variation of the space-time curvature itself.

In conventional inflationary scenarios \cite{GR3,GR4} (see also \cite{BB8} for a general introduction) the maximal frequency of the relic gravitational radiation (denoted hereunder by $\nu_{max}$) can be (at most) ${\mathcal O}(100)$ MHz and three physical aspects of the concordance paradigm contribute to this conclusion: {\it (i)} the curvature scale reached at the end of inflation, {\it (ii)} the timeline of the post-inflationary expansion and {\it (iii)} the current bounds on the tensor-to-scalar ratio $r_{T}$ already mentioned in the previous paragraph. By altering one of these complementary ingredients the value of $\nu_{max}$ also changes and may even increase above the GHz (as it happens, for instance, in the presence of a long post-inflationary stage stiffer than radiation \cite{GR5}). Another relevant situation is provided by bouncing scenarios\footnote{Bouncing models, originally envisaged by Tolman \cite{BB1} and  Lema\^itre \cite{BB2}, appear in various contexts encompassing the cosmological scenarios inspired by string theory and quantum gravity; see, for instance, \cite{BB3,BB4,BB5} for three reviews with different (and sometimes opposite) viewpoints that however converge, for independent reasons, on the bouncing dynamics. After the bouncing regime the maximal curvature scale at the onset of the decelerated stage of expansion could even be ${\mathcal O}(M_{s})$ (where $M_{s}$ denotes throughout the string mass scale). Typically $M_{s} < M_{P}$ and we shall assume, for the sake of concreteness, $M_{s} = {\mathcal O}(0.01)\, M_{P}$. } where the  maximal curvature scale at the onset of the decelerated stage of expansion may become of the order of the string mass scale $M_{s}$ implying a $\nu_{max}$ that could even exceed the GHz range. This perspective seems incidentally (more?) consistent with the swampland conjectures \cite{BB6,BB7} suggesting that the slow-roll parameters should be ${\mathcal O}(1)$ in contradiction with the very existence of a slow-roll regime. For the sake of concreteness we start the discussion from the Ford-Parker effective action \cite{GR2}
\begin{equation}
S_{g} = \frac{1}{8\ell_{P}^2} \int d^{4} x \sqrt{-\overline{g}}  \, \overline{g}^{\mu\nu} \partial_{\mu}
h_{i\, j} \partial^{\nu} \, h^{i\, j}, \qquad \overline{g}_{\alpha\beta} = a^2(\tau) \, \, \eta_{\alpha\beta}, \qquad \ell_{P} = 1/\overline{M}_{P},
\label{LL1} 
\end{equation}
where we assumed, consistently with the observational data \cite{AD1,AD2},  that the background geometry is conformally flat while $h_{i\, j}$ is both traceless and solenoidal\footnote{The signature of the Minkowski metric is mostly minus (i.e. $[+,\, -,\, -,\, -]$)
and the Greek indices are four-dimensional while the Latin (lowercase) indices are defined on the three-dimensional Euclidian sub-manifold. 
Note that $\overline{M}_{P} = M_{P}/\sqrt{8\,\pi}$; throughout the paper $M_{P}$ and $\overline{M}_{P}$ will be used, depending on the convenience, to simplify some of the equations. Units $\hbar\, =\, c \, =1$ 
will be employed throughout.}; in Eq. (\ref{LL1}) $\overline{g}_{\alpha\beta}$ denotes the background metric while $\tau$ is the conformal time coordinate and $a(\tau)$ indicates the scale factor.  Even though Eq. (\ref{LL1}) is not the most general action compatible with two derivatives of the tensor amplitude, the same analysis can be repeated (with analogous conclusions) in the context of the following generalized action which we mention for the sake of completeness:
 \begin{equation}
 \widetilde{\,S\,}_{g} = \frac{1}{8 \ell_{P}^2} \int d^{3} x\, \int d\tau \biggl[ A(\tau) \partial_{\tau}  h_{i\, j} \partial_{\tau} h^{i\, j} - B(\tau) \,\partial_{k} h_{i\, j} \,\partial^{k} h^{i\, j}\biggr].
 \label{GGq}
 \end{equation}
 Equation (\ref{GGq}) can be brought in its canonical form by introducing a new time coordinate $\eta$ related to the conformal time $\tau$ as $d\tau \sqrt{B(\tau)/A(\tau)} \,=\, d\eta$; thanks to such a coordinate transformation we the have that $ \widetilde{\,S\,}_{g}$
 can be written as:
  \begin{equation}
 \widetilde{\,S\,}_{g} = 
 \frac{1}{8 \ell_{P}^2} \int d^{3} x\, \int d\eta \, b^2(\eta)\, \biggl[ \partial_{\eta}  h_{i\, j} 
 \partial_{\eta} h^{i\, j} -  \partial_{k} h_{i\, j} \,\partial^{k} h^{i\, j}\biggr], \qquad b(\eta) = [A(\eta) \, B(\eta)]^{1/4}.
 \label{GGp}
 \end{equation}
When $A(\tau) = B(\tau) = a^2(\tau)$ we have that $S_{g} = \widetilde{\,S\,}_{g}$ and with the dictionary spelled out by Eqs. (\ref{GGq})--(\ref{GGp}) it is always possible to phrase the conclusions deduced from Eq. (\ref{LL1}) in the concrete situations where the relic gravitons inherit an effective refractive index (possibly generated by more general gravity actions \cite{REF}).

When the quantum mechanical fluctuations dominate against their classical counterpart relic gravitons are produced from the variation of the space-time curvature and this may happen during a stage of accelerated expansion but also in a bouncing stage \cite{REF0}. Even if $\nu_{max}$ could be deduced both classically and quantum mechanically, it is useful to preliminarily associate the tensor amplitudes (and their conformal time derivatives) with the appropriate field operators obeying canonical commutation relations:
\begin{eqnarray}
\widehat{\mu}_{i\,j}(\vec{x}, \tau) &=& \frac{\sqrt{2} \, \ell_{P}}{(2\pi)^{3/2}} \sum_{\lambda= \oplus,\, \otimes} \int 
d^{3} p \, \, e_{i\, j}^{(\lambda)}(\hat{p}) \,\, \widehat{\mu}_{\vec{p}, \, \lambda}(\tau) \,\, e^{- i \vec{p}\cdot\vec{x}},
\nonumber\\
\widehat{\pi}_{i\,j}(\vec{x}, \tau) &=& \frac{1}{4 \sqrt{2} \, \ell_{P} \, (2\pi)^{3/2}} \sum_{\lambda=\oplus,\, \otimes} \int 
d^{3} p \, \, e_{i\, j}^{(\lambda)}(\hat{p}) \,\, \widehat{\pi}_{\vec{p}, \, \lambda}(\tau) \,\, e^{- i \vec{p}\cdot\vec{x}},
\label{LL1a}
\end{eqnarray}
where  $\widehat{\mu}_{i\,j} \, a(\tau) = \widehat{h}_{i\, j}$ and the sum over $\lambda$ runs over the two tensor polarizations defined in the usual manner\footnote{If $\hat{m}$, $\hat{n}$ and $\hat{k}$ are a triplet of mutually orthogonal unit vectors (obeying $\hat{m}\times \hat{n} = \hat{k}$), the two tensor polarizations are defined as  $e^{\oplus}_{i\, j}(\hat{k}) = \hat{m}_{i} \, \hat{m}_{j} - \hat{n}_{i} \, \hat{n}_{j}$ and as  $e^{\otimes}_{i\, j}(\hat{k}) = \hat{m}_{i} \, \hat{n}_{j} + \hat{n}_{i} \, \hat{m}_{j}$.}. Using now Eqs. (\ref{LL1})--(\ref{LL1a}) the Hamiltonian operator becomes
\begin{equation}
\widehat{H}(\tau) = \frac{1}{2} \int d^{3} p \sum_{\lambda=\oplus,\otimes} \biggl\{ \widehat{\pi}_{-\vec{p},\,\lambda}\, \widehat{\pi}_{\vec{p},\,\lambda} + p^2 \widehat{\mu}_{-\vec{p},\,\lambda}\, \widehat{\mu}_{\vec{p},\,\lambda} 
+ {\mathcal H} \biggl[ \widehat{\pi}_{-\vec{p},\,\lambda}\, \widehat{\mu}_{\vec{p},\,\lambda} + 
\widehat{\mu}_{-\vec{p},\,\lambda}\, \widehat{\pi}_{\vec{p},\,\lambda} \biggr] \biggr\},
\label{LL1b}
\end{equation}
and recalling the explicit form of the commutation relations (i.e.  $[\widehat{\mu}_{\vec{k},\, \alpha}(\tau), \widehat{\pi}_{\vec{p},\, \beta}(\tau)] = 
i \, \delta_{\alpha\beta} \, \delta^{(3)}(\vec{k} + \vec{p})$) the evolution of the field operators in the Heisenberg description follows directly from Eq. (\ref{LL1b}) 
\begin{equation}
\widehat{\mu}_{\vec{k}, \, \lambda}^{\prime} = \widehat{\pi}_{\vec{k},\,\lambda} + {\mathcal H} \,\,\widehat{\mu}_{\vec{k}, \, \lambda}, \qquad \widehat{\pi}_{\vec{k}, \, \lambda}^{\prime} = - k^2 \widehat{\mu}_{\vec{k},\,\lambda} - {\mathcal H}\,\, \widehat{\pi}_{\vec{k}, \, \lambda},
\label{LL1c}
\end{equation}
where the prime denotes, as usual, a derivation with respect to the cosmic time coordinate $\tau$;
the expansion rate ${\mathcal H}$ (defined with respect to the conformal time coordinate) and the Hubble rate $H$ are connected as ${\mathcal H} = a\, H= a^{\prime}/a$. Taking now into account Eqs. (\ref{GGq})--(\ref{GGp}), the 
results of Eq. (\ref{LL1c}) can be generalized by formally replacing $\tau\,\to\,\eta$ coordinate and  $a(\tau) \to b(\eta)$. Equation  (\ref{LL1c}) is also equivalent to the following pair of equations:
\begin{equation}
\widehat{\mu}_{\vec{k}, \, \lambda}^{\prime\prime} + \biggl[ k^2 - \frac{a^{\prime\prime}}{a} \biggr] \widehat{\mu}_{\vec{k}, \, \lambda}=0, \qquad \widehat{\pi}_{\vec{k},\,\lambda} = \widehat{\mu}_{\vec{k}, \, \lambda}^{\prime} - {\mathcal H} \,\,\widehat{\mu}_{\vec{k}, \, \lambda}.
\label{LL1d}
\end{equation}
If we would now (formally) replace $a(\tau) \to b(\eta)$ and $\tau \to \eta$ we would obtain the evolution of the field operators derived from the classical action $\widetilde{\,S\,}_{g}$. All in all Eqs. (\ref{LL1c})--(\ref{LL1d}) account for the evolution of the field operators in the Heisenberg representation and, by definition, they coincide with the ones obeyed by the classical amplitudes.

On a classical ground the maximal frequency of the spectrum corresponds to the minimally amplified wavenumber and this conclusion follows from Eq. (\ref{LL1d}) where the wavenumbers $k \simeq | a^{\prime\prime}/a |$ are comparatively less amplified than the ones satisfying\footnote{If the pivotal action is given by Eqs. (\ref{GGq})--(\ref{GGp}) the conditions on the mode $k$ can be expressed in the $\eta$-time parametrization as $k^2 \simeq | b^{\prime\prime}/b |$ and as $k^2 \ll | b^{\prime\prime}/b |$ where the prime would now denote a derivation with respect to the $\eta$-time parametrization. } $k^2 < | a^{\prime\prime}/a |$.  For a more explicit relation we may recall
 that $a^{\prime\prime}/a$ is proportional to the Ricci scalar of the underlying 
 background geometry, i.e. $a^{\prime\prime}/a = {\mathcal H}^2 + {\mathcal H}^{\prime} = a^2 H^2 (2 - \epsilon)$,
where $ \epsilon = - \dot{H}/H^2$ is the standard slow-roll parameter that is 
very small during an accelerated stage of expansion (i.e. $\epsilon \ll 1$)
while $\epsilon_{f} = \epsilon(\tau_{f}) = {\mathcal O}(1)$ in the  
final stages of inflation. For this reason the maximal wavenumber of the spectrum is roughly proportional to $a\,H$ evaluated at the end of the inflationary stage and the field operators of Eq. (\ref{LL1d}) experience their minimal amplification for 
\begin{equation}
k = {\mathcal O}(k_{max}) = a_{f} H_{f} \biggl[ 1  + {\mathcal O}(\epsilon_{f})\biggr], \qquad\qquad H_{f} = {\mathcal O}(H),
\label{LL1f}
\end{equation}
where $H_{f}$ designates the final value of the Hubble rate (i.e. when the accelerated expansion ceases and $\epsilon_{f} = {\mathcal O}(1)$) while $H$ stands for the expansion rate during inflation. Since the expansion rate is nearly constant during a slow-roll stage \cite{BB8}, $H_{f}$ and $H$ are approximately of the same order. The gauge invariant power spectrum of the curvature inhomogeneities generally depends on $H(\tau)$
\begin{equation}
{\mathcal P}_{{\mathcal R}}(k,\tau) = \frac{|k\, \tau|^2}{\pi \, \epsilon(\tau)} \, \frac{H^2(\tau)}{M_{P}^2},
\label{LL1fa}
\end{equation}
but when a given wavelength crosses the comoving Hubble radius during inflation (i.e. $k \tau_{k} = {\mathcal O}(1)$) Eq. (\ref{LL1fa})  becomes:
\begin{equation}
{\mathcal P}(k, \tau_{k}) = {\mathcal P}(k, 1/k) = \frac{1}{\pi\, \epsilon_{k}} \biggl(\frac{H_{k}}{M_{P}}\biggr)^2,\qquad  H_{k}= H(\tau_{k}).
\label{LL1g}
\end{equation}
 For the scales that determine the amplitudes of the 
 temperature and of the polarization anisotropies of the cosmic microwave background, the scalar power spectrum is customarily parametrized as ${\mathcal P}_{{\mathcal R}}(k) = {\mathcal A}_{{\mathcal R}} \, (k/k_{p})^{n_{s}-1}$ where $k_{p} = 0.002\, \mathrm{Mpc}^{-1}$ denotes a conventional pivot scale and $n_{s}$ represents the scalar spectral index \cite{AD1,AD2}. Thus, for scales comparable with $k_{p}$, the value of $H_{k}$ is in fact fixed and it depends both on ${\mathcal A}_{{\mathcal R}}$ and on the tensor-to-scalar ratio:
 \begin{equation}
H_{k}/M_{P} \simeq \sqrt{\pi\, \epsilon_{k} \, {\mathcal A}_{{\mathcal R}}} = \sqrt{ \pi \, r_{T} \, {\mathcal A}_{{\mathcal R}}}/4 \simeq H_{d}/M_{P},
 \label{LL1h}
 \end{equation}
where the second equality is dictated by the consistency relations. Since between $H_{k}$ and $H_{f}$ the background inflates (and the Hubble rate mildly decreases), the 
 value of $k_{max}$ appearing in Eq. (\ref{LL1f}) is ultimately determined by  
 the curvature scale reached at the end of inflation which we identify with the 
 expansion rate at the onset of the decelerated stage of expansion (indicated by
 $H_{d}$ in Eq. (\ref{LL1h}) and in the following part of the discussion).
 
Within the classical perspective of the previous paragraph the maximal frequency $\nu_{max}$ is obtained by redshifting  $k_{max}/(2 \pi)$ from the end of the inflationary stage down to the present epoch. Because in the concordance 
scenario the Universe always expands, the physical frequencies are suppressed 
between the end of inflation and the current time. For this reason the present value of the scale factor is normalized as $a_{0} = 1$ so that comoving and physical frequencies coincide today (but not in the past). It turns then out that the maximal frequency is always subjected to the following upper bound that depends not only on the expansion rate at the onset of the decelerated stage (i.e. $H_{d}$ in Eq. (\ref{LL1h})) but also on the subsequent expansion history
\begin{equation}
\nu_{max} \leq \,\, {\mathcal Q}(g_{s}, g_{\rho}, \Omega_{R0}) \,\,\sqrt{H_{0} \, H_{d}} \, \, \prod_{i=1}^{n-1} \,\, \xi_{i}^{\alpha_{i}},
\label{GG1}
\end{equation}
where $H_{d}$ and $H_{0}$ denote, respectively, the Hubble rate at the onset of the decelerated stage and the present time. The last term appearing at the right-hand side of Eq. (\ref{GG1}) contains a product of different factors accounting for the post-inflationary evolution prior to radiation-dominance and the various $\xi_{i}$ measure the duration of each post-inflationary stage :
\begin{equation}
\xi_{i} = H_{i+1}/H_{i} < 1, \qquad  \alpha_{i}\, =\, (\delta_{i} -1)/[2\, (\delta_{i} + 1)].
\label{GG3}
\end{equation}
The Hubble rate during the {\em i-th} stage is indicated by $H_{i}$ and 
we remark, in this respect, that $\xi_{n-1} = H_{n}/H_{n-1}$ so that the expansion rate $H_{n}$ coincides with the onset of the radiation dominated phase (i.e. $H_{n}= H_{r}$) since we conventionally decided to identify  $a_{n}$  with $a_{r}$, i.e. the scale factor at radiation dominance. Concerning the results of Eq. (\ref{GG1}) two relevant comments involve, respectively, the expansion histories after and before radiation dominance.
\begin{itemize}
\item{} After radiation dominance, in local thermal equilibrium, the entropy density is conserved and the total energy density depends on $g_{\rho}$ (i.e. the number of relativistic degrees of freedom in the plasma) while $g_{s}$ denotes the effective number of relativistic degrees of freedom appearing in the entropy density.  Denoting with $\Omega_{R0}= {\mathcal O}(10^{-5})$ the present critical fraction of relativistic species,
the term ${\mathcal Q}(g_{s}, g_{\rho}, \Omega_{R0})$ of Eq. (\ref{GG1}) can be written as\footnote{In the standard model of particle interactions we could conventionally assume $g_{s,\, r}= g_{\rho,\, r} = 106.75$ and $g_{s,\, eq}= g_{\rho,\, eq} = 3.94$ so that ${\mathcal Q}(g_{s}, \, g_{\rho}, \Omega_{R0})\leq 1.13 \times 10^{-2}$. It seems therefore reasonable to estimate that, at most, ${\mathcal Q}(g_{s}, \, g_{\rho}, \Omega_{R0}) = {\mathcal O}(10^{-2})$. }
\begin{equation}
{\mathcal Q}(g_{s}, g_{\rho}, \Omega_{R0}) = \frac{ (2 \Omega_{R0})^{1/4}}{2 \pi} \,\, \biggl(\frac{g_{s,\, eq}}{g_{s, \, r}}\biggr)^{1/3} \,\biggl(\frac{g_{\rho,\, eq}}{g_{\rho, \, r}}\biggr)^{1/4}.
\label{GG2}
\end{equation}
\item{} Before the dominance of radiation, the product of the different contributions appearing in Eq. (\ref{GG1}) arises in the generic case when between $H_{d}$ and $H_{bbn}$ there are $n$ different expanding stages (not necessarily coinciding with radiation). In Eq. (\ref{GG3}) the standard radiation dominated stage extending from $H_{d}$ down to matter-radiation equality is obtained when $\delta_{i}\to 1$ (and $\alpha_{i} \to 1/2$) for all the various $\delta_{i}$ and $\alpha_{i}$ characterizing the different expanding stages between 
$H_{d}$ and $H_{r}$.
\end{itemize}

A more explicit form of the limit is obtained by maximizing the different contributions at the right-hand side of Eq. (\ref{GG1}) in a way that is compatible with the current cosmological bounds. So, for instance, according to Eq. (\ref{LL1h}),  in conventional inflationary models we typically have\footnote{This upper limit comes about since ${\mathcal A}_{{\mathcal R}} = {\mathcal O} (10^{-9})$ and, as already mentioned, $r_{T} \leq 0.06$ \cite{AD1,AD2}.} $H_{k} \simeq H_{d} \leq {\mathcal O}(10^{-6}) M_{P}$ but the value of $H_{d}$ may be larger in the case of bouncing scenarios where $H_{d}$ can even reach string mass scale $M_{s} = {\mathcal O}(10^{-2}) M_{P}$. Moreover, for $H< H_{d}$ the contributions coming from the expansion rates (and discussed in Eqs. (\ref{GG1})--(\ref{GG3})) are maximized as long as\footnote{Since, by definition, $\xi_{i} <1$, according to Eq. (\ref{GG3}), the different terms acquire the largest value when $\alpha_{i} < 0$ or, which is the same, when $\delta_{i} <1$.} $\delta_{i}<1$ during the various expanding stages. The bound obtained in this manner is further improved  by requiring that the $\alpha_{i}$ are as negative as possible (or, which is the same, that $|\alpha_{i}|$ is maximal). This happens when all the $\delta_{i}$ coincide with a certain  $\overline{\delta} = \delta_{min}$ denoting the absolute minimum of the post-inflationary expansion rate and, in this case, $\nu_{max}$ ultimately depends on $\overline{\xi} = H_{r}/H_{d}$ since, by definition 
 \begin{equation}
\overline{\xi}= \xi_{1} \, \xi_{2} \, .\, .\,.\, \xi_{n-2} \, \xi_{n-1}= H_{r}/H_{d}, \qquad \delta_{1} = \delta_{2}= \,.\,.\,.=\delta_{n-2} = \delta_{n-1} = \overline{\delta} < 1.
\label{GG4}
\end{equation}
In other words, because all the $\delta_{i}$ are equal we obtain that the product of all the $\xi_{i}$ (denoted by $\overline{\xi}$ in Eq. (\ref{GG4}))  is raised to the same power implying that the contribution of the whole decelerated stage of expansion of Eq. (\ref{GG1}) is maximized by a single expanding stage characterized by $\overline{\delta} < 1$:
\begin{equation}
\prod_{i=1}^{n-1} \,\, \xi_{i}^{\alpha_{i}} \leq \biggl(\frac{H_{r}}{H_{d}}\biggr)^{ \overline{\alpha}} < \biggl(\frac{H_{bbn}}{H_{d}}\biggr)^{\overline{\alpha}}, 
\qquad \overline{\xi} < 1, \qquad \overline{\alpha} = \frac{\overline{\delta}-1}{2 (\overline{\delta} +1)} < 0.
\label{GG5}
\end{equation}
The chain of inequalities in Eq. (\ref{GG5}) is derived from Eq. (\ref{GG4}) because $\overline{\xi}$ ultimately coincides with $H_{r}/H_{d}$ where, as already mentioned, $H_{r}$ estimates the curvature scale of radiation dominance. The lowest $H_{r}$ coincides with the curvature scale of big-bang nucleosynthesis typically ranging between $10^{-42}\, M_{P}$ and $10^{-44} \, M_{P}$; therefore the relevant upper limit is obtained, as indicated in Eq. (\ref{GG5}), by setting $H_{r} \to H_{bbn}$. 

An explicit value of $\overline{\delta}=\delta_{min}$ (or, equivalently, of $\overline{\alpha}$) would make the bound of Eq. (\ref{GG1}) even more transparent. Let us then consider, for this purpose, a perfect fluid and note that the maximal value of the barotropic index (be it $w_{max}$) corresponds to the minimal value of the expansion rate, i.e.  $\delta_{min} = 2/(3 w_{max} +1)$. But at most $w_{max} \to 1$ and in this situation the sound speed of the plasma coincides with the speed of light; thus we have that $\overline{\alpha} = \alpha_{min} \to -1/6$. Along a slightly different viewpoint the value of $\delta_{min}$ is determined by a stage where the 
expansion is governed by an oscillating scalar field $\varphi$ but, in this case, the value of $\overline{\alpha}$ is in fact the same\footnote{If the minimum of the potential is located in $\varphi =0$, for $ \varphi < M_{P}$ we can parametrize $W(\varphi)$ as $W_{1} \Phi^{2 q}$ (with $\Phi = \varphi/M_{P}$). The coherent oscillations of the inflaton imply that the energy density of the scalar field is roughly constant and, in average, the expansion rate is $\delta=  (q+1)/(2 q-1)$ \cite{CC1}. The value of $\alpha$ becomes then $2(1-q)/(6 q +1)$ so that $ \overline{\alpha} = \alpha_{min} = -1/6$ and  this figure corresponds to the regime where $ q \gg 1$ and $\overline{\delta} \to 1/2$.}. Thanks to these observations, as anticipated, the bound of Eqs. (\ref{GG1}) and (\ref{GG5}) is made more manifest and its general form is:
\begin{equation}
\nu_{max} < 10^{6} \, \biggl(\frac{H_{d}}{M_{P}}\biggr)^{2/3}\,\, \biggl(\frac{h_{0}^2 \Omega_{R0}}{4.15\times 10^{-5}}\biggr)^{1/4} \, \, \mathrm{THz}.
\label{GG6}
\end{equation}
 In a conservative perspective we assumed $H_{r} \to H_{bbn} = 10^{-42} \, M_{P}$ in 
Eq. (\ref{GG6}); this implies that the background was dominated by radiation already for temperatures ${\mathcal O}(10)$ MeV. In the context of conventional inflationary models $H_{d} = {\mathcal O}(H_{k}) = {\mathcal O}(10^{-6}) \, M_{P}$ and this means that the bound of Eq. (\ref{GG6}) implies $ \nu_{max} < {\mathcal O}(100)\, \mathrm{THz}$. In the case of bouncing scenarios we have instead that $H_{d} \leq M_{s}$ implying $\nu_{max} < {\mathcal O}(10^{4})\, \mathrm{THz}$.
 
Equation (\ref{GG6}) follows when the minimally amplified mode is identified with the maximal frequency of the spectrum but a more stringent limit stems directly from the quantum mechanical analysis of the parametric amplification described by Eq. (\ref{LL1b}). If the field operators appearing in Eq. (\ref{LL1b}) are expressed in terms of the corresponding creation and annihilation operators associated with gravitons of opposite three-momenta we have
\begin{equation}
\widehat{\mu}_{\vec{k},\, \lambda} = (\widehat{a}_{\vec{k},\, \lambda} + \widehat{a}_{-\vec{k},\, \lambda}^{\dagger})/\sqrt{2\, k}, \qquad \widehat{\pi}_{\vec{k},\, \lambda} = -  i\, (\widehat{a}_{\vec{k},\, \lambda} - \widehat{a}_{-\vec{k},\, \lambda}^{\dagger})\sqrt{k/2}, 
\label{GG7}
\end{equation}
with $[ \widehat{a}_{\vec{k},\, \lambda}, \widehat{a}_{\vec{p},\, \lambda^{\prime}}] = \delta^{(3)}(\vec{k} - \vec{p})\, \delta_{\lambda\lambda^{\prime}}$. The time evolution of $\widehat{a}_{\vec{p},\,\lambda}(\tau)$ and $\widehat{a}_{-\vec{p},\,\lambda}^{\dagger}(\tau)$ is usually described  in terms of two complex functions $u_{p,\,\lambda}(\tau)$ and $v_{p,\,\lambda}(\tau)$
 \begin{eqnarray}
\widehat{a}_{\vec{p},\,\lambda}(\tau) &=& u_{p,\,\lambda}(\tau)\,\, \widehat{b}_{\vec{p},\,\lambda}-  
v_{p,\,\lambda}(\tau)\,\, \widehat{b}_{-\vec{p},\,\lambda}^{\dagger},  
\label{NT9}\\
\widehat{a}_{-\vec{p},\,\lambda}^{\dagger}(\tau) &=& u_{p,\,\lambda}^{\ast}(\tau) \,\,\widehat{b}_{-\vec{p},\,\lambda}^{\dagger}  -  v_{p,\,\alpha}^{\ast}(\tau)\,\, \widehat{b}_{\vec{p},\,\lambda},
\label{NT10}
\end{eqnarray}
where $| u_{p,\,\lambda}(\tau)|^2 - | v_{p,\,\lambda}(\tau)|^2 =1$ because the whole transformation 
must be unitary to preserve the form of the commutation relations 
between the field operators. If Eqs. (\ref{NT9})--(\ref{NT10}) are now inserted into Eq. (\ref{LL1c}), the evolution of $u_{p,\,\lambda}(\tau)$ and $v_{p,\,\lambda}(\tau)$ is given by:
\begin{equation}
u_{p,\,\lambda}^{\prime} = - i p\, u_{p,\,\lambda}  - {\mathcal H}\,  v_{p,\,\lambda}^{\ast}, 
\qquad v_{p,\,\lambda}^{\prime} = - i p\, v_{p,\,\lambda} -{\mathcal H} \,\,u_{p,\,\lambda}^{\ast},
\label{NT11}
\end{equation}
The averaged multiplicity is obtained by computing the mean number of gravitons for with momentum $\vec{k}$ and $- \vec{k}$, i.e. 
\begin{equation}
\langle \hat{N}_{k} \rangle = \sum_{\lambda=\oplus, \, \otimes} \, \langle \widehat{a}_{\vec{k},\,\lambda}^{\dagger} \widehat{a}_{\vec{k},\, \lambda} + \widehat{a}_{-\vec{k},\,\lambda}^{\dagger} \widehat{a}_{-\vec{k},\,\lambda} \rangle = 4 \, \,\overline{n}(k,\tau),
\label{NT12}
\end{equation}
where $\overline{n}(k,\tau)= |v_{k}(\tau)|^2$ denotes the multiplicity of the pairs of relic gravitons and the further factor of $2$ counts the polarizations. From Eq. (\ref{NT12}) the spectral energy density in critical units is expressed in terms of the averaged multiplicity of the produced gravitons with opposite three-momenta as: 
\begin{equation}
\Omega_{gw}(\nu, \tau_{0}) = \frac{128\, \pi^3}{3} \,\, \frac{\nu^{4}}{H_{0}^2 \, M_{P}^2}\,\, \overline{n}(\nu, \tau_{0}).
\label{HH1}
\end{equation}
According to Eq. (\ref{HH1}) the maximal frequency corresponds to the situation where a single pair relic gravitons is produced; this single graviton limit (or single graviton line) \cite{DD1} is realized when $\overline{n}(\nu_{max}, \tau_{0}) \to 1$. In other words, the minimally amplified wavenumber introduced in Eq. (\ref{LL1f}) coincides, in a quantum mechanical interpretation, with the production of a single graviton pair. This observation
is very similar to the one originally discussed by Dyson \cite{FA} with the difference 
that, in the present context, the nature of the production process is fixed 
by the evolution of the space-time curvature \cite{DD1}.  All the wavelengths reentering the Hubble radius between the end of inflation and BBN must notoriously comply with the bound\footnote{Equation (\ref{HH2}) sets an indirect constraint  on the extra-relativistic species possibly present at the time of nucleosynthesis \cite{DD2,DD3,DD4}. Since Eq. (\ref{HH2}) is also relevant in the context of neutrino physics the limit is  often expressed in terms of $\Delta N_{\nu}$ (i.e. the contribution of supplementary neutrino species). The actual bounds on $\Delta N_{\nu}$ range from $\Delta N_{\nu} \leq 0.2$ to $\Delta N_{\nu} \leq 1$ so that the integrated spectral density in Eq. (\ref{HH2}) must range, at most, between  $10^{-6}$ and $10^{-5}$.} 
\begin{equation}
h_{0}^2 \, \int_{\nu_{bbn}}^{\nu_{max}} \,\Omega_{gw}(\nu,\tau_{0}) \,\,d\ln{\nu} < 5.61\times 10^{-6} \biggl(\frac{h_{0}^2 \,\Omega_{\gamma0}}{2.47 \times 10^{-5}}\biggr) \, \Delta N_{\nu},
\label{HH2}
\end{equation}
where $\Omega_{\gamma 0}$ is the (present) critical fraction of CMB photons. 
In a different perspective the diffuse background of relic gravitons was produced 
{\em after} BBN. In this case we observe that $\Omega_{gw}(\nu, \tau_{0})$ 
must always be smaller than the the current fraction of relativistic species to avoid an observable impact on the CMB and matter power spectra \cite{DD5,DD6}. The two bounds are actually numerically compatible so that the final limit on $\nu_{max}$ is given by:
\begin{equation}
\nu_{max} \leq 0.165 \,\, \Omega_{R0}^{1/4}\,\, \sqrt{H_{0}\, M_{P}} < \,\mathrm{THz},
\label{HH3}
\end{equation}
where the inequality follows in the limit $\overline{n}(\nu_{max}, \tau_{0}) \to {\mathcal O}(1)$, i.e. in the case
where a single pair of gravitons is produced. The bound of Eq. (\ref{HH3}) is numerically more restrictive 
than the one of Eq. (\ref{GG6}) even if the two are compatible. While Eq. (\ref{GG6}) 
is derived from a series of classical considerations, Eq. (\ref{HH3}) stems more directly 
from the quantum description of the parametric amplification. 

Up to now we demonstrated that the cosmic gravitons produced in conventional and unconventional inflationary scenarios must satisfy the quantum bound of Eq. (\ref{HH3}) and, a fortiori, the classical bound of Eq. (\ref{GG6}). It is now useful to  verify this conclusion by considering scenarios based on different physical premises. When a conventional inflationary stage is followed by a radiation epoch extending down to the scale of matter-radiation equality we 
the maximal frequency is
\begin{equation}
\overline{\nu}_{max} =206.52\, \biggl(\frac{{\mathcal A}_{{\mathcal R}}}{2.41\times 10^{-9}}\biggr)^{1/4}\,\,
\biggl(\frac{r_{T}}{0.06}\biggr)^{1/4} \,\, \biggl(\frac{h_{0}^2 \, \Omega_{R\,0}}{4.15\times 10^{-5}}\biggr)^{1/4} \,\,\mathrm{MHz},
\label{HH4}
\end{equation}
so that the upper bound of Eq. (\ref{HH3}) is satisfied.  A further possibility is to complement a conventional inflationary phase with a 
long post-inflationary stage expanding slower than radiation; in this case 
the largest possible value of the maximal frequency is obtained 
when $\overline{\delta} \to 1/2$ and if $H_{r}$ coincides with the electroweak curvature scale $H_{ew}$ we obtain:
\begin{equation}
\nu_{max} = 94.6 \biggl(\frac{H_{r}}{H_{ew}}\biggr)^{1/6} \, \mathrm{GHz} < \mathrm{THz},
\label{HH5}
\end{equation}
where we took $H_{ew} = 10^{-32} \, M_{P}$. If the post-inflationary phase 
is faster than radiation (i.e. $\overline{\delta}> 1$) the maximal frequency 
is smaller than $200$ MHz and the bound of Eq. (\ref{HH3}) is satisfied to a greater acuracy. We  consider next a bouncing scenario where the post-inflationary stage of expansion is dominated by radiation; in this case $\nu_{max}$ can be estimated as:
\begin{equation}
\nu_{max} = 0.1 \, \biggl(\frac{H_{d}}{M_{P}}\biggr)^{1/2} \,\, \mathrm{THz} < \mathrm{THz}, 
\label{HH6}
\end{equation}
where $H_{d} < M_{P}$. If $H_{d}$ coincides with the string scale we would 
have approximately that $H_{d} = M_{s} = {\mathcal O}(10^{-2})\, \mathrm{M}_{P}$ 
so that $\nu_{max}$ may range between $10$ and $100$ GHz.   As a last 
example we consider the thermal graviton spectrum \cite{DD1}. 
The temperature of the gravitons is always smaller than the one of the photons; in particular we 
denoting with $T_{g\, 0} $ and $ T_{\gamma\,0}$ the current values of the graviton and photon 
temperatures they are related as \cite{DD1}
\begin{equation}
T_{g\, 0} = \epsilon_{g} \,\, T_{\gamma\,0} \leq 0.9051\,\, \biggl(\frac{T_{\gamma\, 0}}{2.72548\, \mathrm{K}}\biggr)\,\,\, \mathrm{K},
\end{equation}
where $\epsilon_{g}$ depends on the number of relativistic species at the time of graviton decoupling (close to the Planck epoch). The maximal frequency of the thermal black-body spectrum is then 
\begin{equation}
\nu_{max} = 73.943 \biggl(\frac{T_{g\, 0}}{0.9051 \, \mathrm{K}}\biggr) \, \mathrm{GHz}< \mathrm{THz},
\label{HH7}
\end{equation}
corresponding to a spectral energy density in critical units $h_{0}^2 \Omega_{gw}(\nu_{max}, \tau_{0}) \leq 2.213\times 10^{-7}$ as long as $\epsilon_{g} \leq 0.3321$. The examples of  Eqs. (\ref{HH5})--(\ref{HH6}) and (\ref{HH7}) demonstrate not only that the bound of Eq. (\ref{HH3}) 
is always satisfied in concrete situations but also that $\nu_{max}$ is numerically 
similar in different scenarios based on diverse physical premises.

Since the bound of Eq. (\ref{HH3}) does not depend on the specific model it is practical to normalize the potential chirp amplitude directly in the THz domain\footnote{The spectral energy density at the present 
time and the chirp amplitude $h_{c}(\nu, \tau_{0})$ are 
related as $\Omega_{gw}(\nu,\, \tau_{0}) = 2\, \pi^2 \, \nu^2 h_{c}^2/( 3 \, H_{0}^2)$.} where 
the bound on the maximal frequency can be converted into a limit on $h_{c}$.
If the spectral energy is normalized in the THz domain with a certain high-frequency 
slope  $\nu^{m_{T}}$, Eqs. (\ref{HH1})--(\ref{HH2}) imply that the minimal chirp amplitude required for the direct detection of cosmic gravitons 
should be smaller than ${\mathcal O}(10^{-32})$ or, more accurately:
\begin{equation}
h^{(min)}_{c}(\nu, \tau_{0}) < 8.13 \times 10^{-32} \biggl(\frac{\nu}{0.1\, \mathrm{THz}}\biggr)^{-1 +m_{T}/2}.
\label{MM1}
\end{equation}
 The negative aspect of Eq. (\ref{MM1}), as already stressed before in a model-dependent perspective \cite{DD1}, is that a sensitivity ${\mathcal O}(10^{-20})$ or even ${\mathcal O}(10^{-24})$ in the chirp amplitude for frequencies in the MHz or GHz regions is irrelevant for a direct or indirect detection of high-frequency gravitons. It has been actually suggested long ago that microwave cavities \cite{EE1,EE2,EE3,EE4} operating in the MHz and GHz regions could be used for the detection of relic gravitons \cite{GR5}. The same class of instruments has been 
also invoked in \cite{FA1,FA2,FA3,FA4} with the difference that, unlike previous studies (more aware of the potential sources and of the instrumental noises), the required chirp amplitudes are now optimistically set in the range $h_{c}^{(min)} = {\mathcal O}(10^{-20})$ for arbitrarily high frequencies\footnote{Clearly  $h^{(min)}_{c} = {\mathcal O}(10^{-20})$ is an interesting technological achievement that is however more than $10$ orders magnitude larger than the sensitivities required for the detection of cosmic gravitons.}. The present considerations clarify, in a model-independent perspective, that $h_{c}^{(min)}$ must be at least ${\mathcal O}(10^{-32})$ (or smaller) for a potential detection of cosmic gravitons in the THz domain.  There is, however, also a more positive aspect related to the result of Eq. (\ref{MM1}): while for $m_{T} > 2$ the largest signal occurs at the largest frequency, for $m_{T} \leq 2$ frequencies smaller than the THz are experimentally more convenient. Let us consider, for instance, the case $m_{T} \to 1$ which is, incidentally, typical of a post-inflationary 
stiff phase;  we neglect here all the possible logarithmic corrections arising 
in the spectrum even if they sightly change the final estimates \cite{GR5}. In this situation we would have that the chirp amplitude at smaller  frequencies (say in the MHz range) could be ${\mathcal O}(10^{-28})$ (as also proposed in Refs. \cite{EE3,EE4} on the basis of more experimental considerations). In the case $m_{T} \to 2$ (typical of the ekpyrotic scenario) we would have instead that $h_{c}(\nu,\tau_{0})$ is the same at higher and smaller frequencies  \cite{EK1,EK2}. 
Finally for $m_{T} \to 3$ (as it happens in the case of the pre-big bang scenario \cite{EK3}) the chirp amplitude at lower frequencies gets even smaller. There is therefore 
a trade-off between the optimal frequency, the features of the signal and the 
noises (especially the thermal one) indicating that the highest possible frequency 
(close to $\nu_{max}$) is not always the most convenient. Also this aspect should be 
taken into account if the goal is really an accurate assessment of the required sensitivities of high-frequency instruments.

The limits following from Eqs. (\ref{HH3}) and (\ref{MM1}) are also important for the analysis of the statistical properties of the relic gravitons and, in particular, of their degrees of first- and second-order coherence. These observables follow by generalizing the appropriate Glauber correlators \cite{HBT0a,HBT0b} to the expectation values of tensor fields (see Refs. \cite{HBT1} and discussions therein); besides the physical aspects (discussed over a decade ago \cite{HBT2}) the main technical difference between the gravitons and the photons involves the polarization structure of the correlation functions. Mutatis mutandis the physical idea is however similar: if cosmic gravitons are detected by independent interferometers the correlated outputs are employed to estimate the degrees of second-order coherence. The analysis of the interplay between the Hanbury-Brown Twiss (HBT) interferometry and the high-frequency gravitons has been recently discussed in Ref. \cite{DD1} (see also \cite{HBT1,HBT2}); for the present purposes we avoid the polarization dependence appearing in Eqs. (\ref{NT9})--(\ref{NT10}) and introduce 
 the {\em single-particle} (inclusive) density \cite{HBT3a,HBT3b}
\begin{equation}
\rho_{1}(\vec{k}) = \langle \widehat{A}^{\,\dagger}(\vec{k})\,\,\widehat{A}(\vec{k})\rangle, \qquad \widehat{A}(\vec{k}) 
=\int d^{3} p \,\, \widehat{a}_{\vec{p}} \,\, {\mathcal W} (\vec{k} - \vec{p}),
\label{HH1a}
\end{equation}
where $\widehat{A}(\vec{k},\tau)$ (and its Hermitian conjugate) are just a set of creation and annihilation operators that are non-zero inside the volume of the particle source associated with the three-dimensional integral (in real space)
of an appropriate window function ${\mathcal W}(\vec{x})$. By definition, $[\widehat{A}(\vec{k}),\,\,\widehat{A}^{\dagger}(\vec{k})] = \int d^{3} x |{\mathcal W}(\vec{x})|^2$. In the theory of Bose-Einstein interference \cite{HBT3a,HBT3b}
 \begin{equation}
 \rho_{2}(\vec{k}_{1}, \vec{k}_{2}) = \langle \widehat{A}^{\,\dagger}(\vec{k}_{1})\,\,\widehat{A}^{\,\dagger}(\vec{k}_{2})\,\,\widehat{A}(\vec{k}_{2})\,\,\widehat{A}(\vec{k}_{1})\rangle,
 \label{HH2a}
 \end{equation}
 is the {\em two-particle} inclusive density and according to Eqs. (\ref{HH1a})--(\ref{HH2a}) the normalized second-order correlation function
 \begin{equation}
 C_{2}(\vec{k}_{1}, \vec{k}_{2}) = \frac{\rho_{2}(\vec{k}_{1}, \vec{k}_{2})}{\rho_{1}(\vec{k}_{1})\,\,\rho_{1}(\vec{k}_{2})} \to 3 + {\mathcal O}\biggl(\frac{1}{\sqrt{\overline{n}(k_{1})\, \overline{n}(k_{2})}}\biggr),
 \label{HH3a}
 \end{equation}
 estimates the degree of second-order coherence \cite{HBT1,HBT2}. The value of $C_{2}(\vec{k}_{1}, \vec{k}_{2})$ is  always enhanced in comparison with so-called Poissonian limit so that the statistics of the relic gravitons is always super-Poissonian and generally super-chaotic. Indeed in the limit of a large number of graviton pairs $C_{2}(\vec{k}_{1}, \vec{k}_{2}) \to 3$ whereas $C_{2}(\vec{k}_{1}, \vec{k}_{2}) \to 2$ in the case of a chaotic mixture. This result is slightly refined by taking into account the polarisation structure of the correlators, as already discussed in the past \cite{HBT1,HBT2}; in this case $C_{2}(\vec{k}_{1}, \vec{k}_{2}) \leq 3$ but the statistics always remains super-Poissonian.  While the the statistical properties of the relic gravitons determine the degrees of first- and second-order coherence, their potential detection depends from the achievable $h_{c}^{(min)}$ which is not the same in different ranges of comoving frequency. It is therefore not surprising that the analyses of the Bose-Einstein correlations overlooking the physical properties of the cosmic gravitons are often ambiguous and ultimately superficial.

All in all we demonstrated, both at the classical and quantum level, that the largest frequency of the relic gravitons never exceeds the THz band while the minimal detectable chirp amplitude should be at least ${\mathcal O}(10^{-32})$ (or smaller) if the (hypothetical) detectors in the THz domain could claim (even in principle) the detection of a relic signal. However, if the pivotal frequencies of the instruments are reduced from the THz to the GHz (or even MHz) band the minimal required chirp amplitude may increase.

The author wishes to acknowledge relevant discussions with the late Ph. Bernard, G. Cocconi and E. Picasso. It is also a pleasure to thank A. Gentil-Beccot, L. Pieper, S. Rohr and S. Reyes of the CERN Scientific Information Service for their kind help during the preparation of this manuscript.

\end{document}